\documentclass[twocolumn]{jpsj3}
\usepackage{bm}
\def\ie{i.e.}

\newcommand{\lsim}
 {\ \raise.35ex\hbox{$<$}\kern-0.75em\lower.5ex\hbox{$\sim$}\ }
\newcommand{\gsim}
 {\ \raise.35ex\hbox{$>$}\kern-0.75em\lower.5ex\hbox{$\sim$}\ }
\title{
Mott Transition and Spin Structures of Spin-1 Bosons in Two-Dimensional Optical Lattice
at Unit Filling
}
\author{
Yuta \surname{Toga}\thanks{E-mail address: toga@solid.apph.tohoku.ac.jp}, 
Hiroki \surname{Tsuchiura},
Makoto \surname{Yamashita}$^{1,2}$, 
Kensuke \surname{Inaba}$^{1,2}$, and 
Hisatoshi \surname{Yokoyama}$^{3}$
}
\inst{
\address{Department of Applied Physics, Tohoku University, Sendai 980-8579, Japan} \\
$^{1}$\address{NTT Basic Research Laboratories, NTT Corporation, Atsugi, Kanagawa 243-0198, Japan } \\
$^{2}$\address{Japan Science and Technology Agency, CREST, Chiyoda, Tokyo 102-0075, Japan}\\
$^{3}$\address{Department of Physics, Tohoku University, Sendai 980-8578, Japan }\\
}
\abst{
We study the ground state properties of spin-1 bosons in a two-dimensional optical lattice, 
by applying a variational Monte Carlo method to the $S=1$ Bose-Hubbard model on a square lattice at unit filling. 
A doublon-holon binding factor introduced in the trial state 
provides a noticeable improvement in 
the variational energy over the conventional Gutzwiller wave function and 
allows us to deal effectively with 
the inter-site correlations of particle densities and spins. 
We systematically show how spin-dependent interactions modify 
the superfluid-Mott insulator transitions in the $S=1$ Bose-Hubbard model 
due to the interplay between the density and spin fluctuations of bosons. 
 Furthermore, regarding the magnetic phases 
in the Mott region, the calculated spin structure factor elucidates the emergence of nematic and ferromagnetic 
spin orders for antiferromagnetic ($U_2>0$) and  ferromagnetic ($U_2<0$) couplings, respectively. 
}
\kword{
Mott transition, superfluid, insulating state, nematic phase, ferromagnetism, 
$S=1$ Bose-Hubbard model, doublon-holon binding, 
variational Monte Carlo method
}
\begin{document}
\maketitle
Recent progress in ultracold atom experiments has offered 
unprecedented opportunities for exploring  
fundamental quantum phenomena in strongly correlated many-body systems that have been 
largely ignored in conventional condensed-matter physics. 
A prominent example of such phenomena is the quantum phase transition from a superfluid (SF) 
to a Mott insulator (MI), demonstrated using cold bosons with frozen spin degrees of freedom 
trapped in optical lattices\cite{greiner}.
Furthermore, quantum gas microscope techniques \cite{bakr1,bakr2,sherson,weitenberg} 
have opened the door to the detection and manipulation of 
single bosons at a single site level in an optical lattice, 
just like scanning tunneling microscopy in solid state physics.
Quite recently, Endres {\it et al.} \cite{endres} used this technique to track the SF-MI transition in more detail, 
and found that correlated pairs consisting of a doubly populated site (doublon, D) and an unpopulated site (holon, H), which represent the excitations
in an MI, fundamentally determine the properties of the SF-MI transition,
which is consistent with recent numerical studies \cite{manuela,yokoB}.

Now theoretical interest is naturally moving towards the quantum phase transitions in bosons
with {\it unfrozen} spin degrees of freedom\cite{gasA,gasB} trapped in optical lattices.  
The doublon, which is one fragment of the elementary excitation in an MI, will have internal spin
structures unlike spinless (spin-frozen) systems.
Thus, we can strongly expect the interplay between spin correlations and the SF-MI transition to play key roles in the ground state of the systems.
The simplest of such systems will be $S=1$ bosons on an optical lattice, whose properties are
 well captured by the $S=1$ Bose-Hubbard model (BHM)\cite{demler}
\begin{eqnarray}
 {\cal H} &=& -t\sum_{\langle i,j\rangle}\sum_{\alpha} 
\left( \hat{a}_{i,\alpha}^{\dagger}\hat{a}_{j,\alpha} 
     + \hat{a}_{j,\alpha}^{\dagger}\hat{a}_{i,\alpha} \right)
                  - \mu\sum_{i}\hat{n}_{i}
\nonumber \\
          & &+ \frac{U_{0}}{2}\sum_{i}\hat{n}_{i}(\hat{n}_{i}-1) 
             + \frac{U_{2}}{2}\sum_{i}\left( \hat{\bm S}_{i}^{2} 
                                             - 2\hat{n}_{i} \right) ,
\label{hamil}
\end{eqnarray}
where $a_{j,\alpha}$ is an annihilation operator of a boson of spin $\alpha$ 
at the site $j$, 
$\hat n_{j} = \sum_\alpha \hat{a}_{j,\alpha}^{\dagger}\hat{a}_{j,\alpha}$ and $t, U_0 > 0$.
Here, $\alpha=-1,0,1$, and $\langle i,j \rangle$ denotes 
a nearest-neighbor-site pair; the definition of $t$ is a half of that reported in some studies.
The spin-dependent (last) term in eq.~(\ref{hamil}) induces spin mixing and enriches the physics 
of this model, compared with spinless models. 
In cold-atom systems, the values of $U_0$ and $U_2$ depend on the 
$s$-wave scattering wavelength characteristic of the atom species; 
$U_2>0$ ($<0$) for Na (Rb) atoms. 
In contrast to more familiar Fermi-Hubbard models with $S=1/2$ where an antiferromagnetic 
superexchange interaction
prevails in the strongly correlated regime, the $S=1$ BHM exhibits complicated effective inter-site spin interactions
that lead to exotic magnetic phases\cite{demlerPRA,snoek}, owing to the absence of the Pauli exclusion principle. 
Thus far, the phase diagram of this model has been studied based on mean-field type theories\cite{krut,tsuchiya} including a Gutzwiller
approximation (GA)\cite{kimura,yama}, and on density matrix renormalization group\cite{rizzi,sara} and quantum Monte Carlo (QMC) methods for the one-dimensional
system\cite{apaja,batrouni}.
Alternatively, at the cost of the density fluctuation, an effective spin Hamiltonian obtained by a strong-coupling expansion
\cite{demlerPRA,tsuchiya,yip}
was studied to explore the magnetic structures in the MI phase using QMC calculations in two and three dimensions\cite{haradaPRB}.
%
%
%
%
%

In this letter, we provide a consistent description of the ground state properties 
and the phase transition from SF to MI 
of the $S = 1$ BHM on a square lattice,
focusing on correlated pair excitations based on a variational Monte Carlo (VMC) approach, 
which is beyond conventional mean-field and GA techniques. 
We consider the simplest case of an SF-MI transition so that 
we restrict the particle density $\rho = N/N_{s}$ ($N$: particle number, $N_{s}$: the total number of sites) to $\rho=1$
(unit filling) and put $\mu = 0$.
We consider the behavior for other odd commensurate densities ($\rho=3,5,\cdots$) 
to be essentially the same.
We will report even-$\rho$ cases separately. 
To implement VMC calculations, we employ an occupation number representation 
at each site, $|n_1, n_0, n_{-1}\rangle$. 
As a variational wave function, we use a Jastrow-type, 
$
|\Psi_{\rm DH}\rangle={\mathcal P}_{\rm DH} {\mathcal P}_{\rm G} |\Phi\rangle 
$. 
Here $|\Phi\rangle$ is the one-body part: 
$
|\Phi\rangle=\frac{1}{\sqrt{N_{\rm s}!}}
\left(\hat a_{{\bf 0},1}^\dag+\hat a_{{\bf 0},0}^\dag
     +\hat a_{{\bf 0},-1}^\dag\right)^{N_{\rm s}}|0\rangle, 
$ 
where, $\hat a_{{\bf 0},1}^\dag$ indicates 
the ${\bm k}={\bf 0}$ component of the Fourier transformation 
of $\hat a_{j,1}^\dag$. 
Since the total $S_{z}$ is well conserved in cold atom systems, 
we work in the subspace of $\sum_{j}S_{j}^{z} = 0$.
%
As an onsite correlation factor, the Gutzwiller projection
is extended so that it depends on the spin configurations in the site, \cite{yama}
\begin{equation}
{\mathcal P}_{\rm G}=\prod_j\gamma(n_{j,1}, n_{j,0}, n_{j,-1})
|n_1, n_0, n_{-1}\rangle_j\ _j\langle n_1,n_0,n_{-1}|, 
\label{Gutz}
\end{equation}
where coefficients $\gamma$ are variational parameters controlling 
$n_{j,\alpha}$, the occupation particle number of spin $\alpha$ on the site $j$.  
The dependence on spin configuration is necessary for $U_2\ne 0$. 
Since we have confirmed that the probability $P(n)$ for $n\ge 4$ becomes negligible for 
the value of interest, $U_{0}/t\gsim 10$, 
we impose a restriction, $\gamma(n_{j,1}, n_{j,0}, n_{j,-1})=0$, 
on the total occupation number ($n$) at each site for 
$n_j\equiv n_{j,1}+n_{j,0}+n_{j,-1}\ge 4$.
%
%
The D-H correlation factor $\mathcal{P}_{\rm DH}(\eta,\eta\prime)$ used here is the same as that introduced in refs. \citen{manuela} and \citen{yokoB},
where $\eta$ ($\eta'$) is a variational parameter ($0\le\eta\le 1$) that controls the strength of the 
D-H binding between nearest-neighbor (lattice-diagonal) sites. 
For $\eta=1$, isolated doublons and holons are prohibited. 
We assume $\eta$ and $\eta'$ are independent of the onsite spin configuration for simplicity.
%
%
It should be noted here that in effect a multiply populated site means a doublon for $U_0/t\gsim20$, because 
$P(n)$ for $n\ge 3$ almost vanishes.

In the VMC calculations, we first optimized the variational parameters 
using a quasi-Newton method, and calculated the quantities for sets of model parameters
 ($U_0/t$, $U_2/t$) with several million configurations for several system sizes 
of $N_s=L\times L$ sites with $L=8$-$24$. 

%
\begin{figure}
\begin{center}
\vskip 2mm 
\includegraphics[clip,width=8cm]{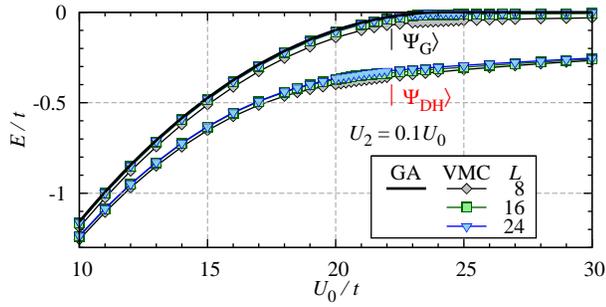}
\end{center}
\vskip -4mm 
\caption{
(Color online) The total energy per site of $|\Psi_{\rm G}\rangle$ and $|\Psi_{\rm DH}\rangle$ 
is compared as a function of $U_0/t$ for three system sizes. 
The VMC data of $|\Psi_{\rm G}\rangle$ for finite $L$'s will converge to the GA result, 
the exact analytic result of $|\Psi_{\rm G}\rangle$ for $L=\infty$. 
The ratio $U_2/U_0$ is fixed at 0.1, and the result is similar to that 
of $U_2=0$. 
}
\vskip -4mm 
\label{eng}
\end{figure}
%
We start by discussing the features of a Mott transition in $|\Psi_{\rm DH}\rangle$  by
comparing those obtained with a Gutzwiller wave function 
$|\Psi_{\rm G}\rangle=\mathcal{P}_{\rm G}|\Phi\rangle$ when the ratio $U_2/U_0$ is fixed at $0.1$.
Figure \ref{eng} shows how the total energy $E/t$ is improved by employing 
 $|\Psi_{\rm DH}\rangle$ in the region of intermediate correlation strength.
The value of $|\Psi_{\rm G}\rangle$ arrives at zero at the Brinkman-Rice transition point\cite{BR}, 
$U_0^{\rm BR}/t\sim 24.3$.
On the other hand, $E/t$ of $|\Psi_{\rm DH}\rangle$ is considerably less than that of GA, 
and remains finite even for large $U_0/t$ values.
\par

\begin{figure}
\begin{center}
\vskip 2mm 
\includegraphics[clip,width=8cm]{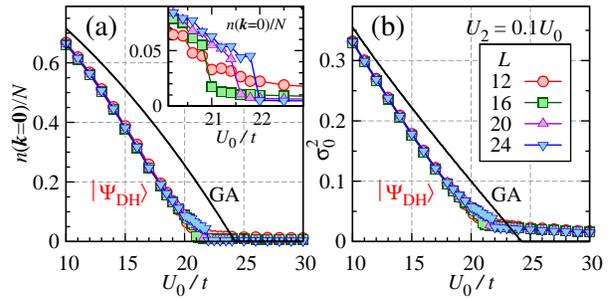}
\end{center}
\vskip -4mm 
\caption{
(Color online) (a) Condensate fraction or momentum distribution function 
at ${\bm k} = (0,0)$ and (b) density fluctuation, as a function of 
$U_0/t$ for $|\Psi_{\rm G}\rangle$ and of $|\Psi_{\rm DH}\rangle$ for $U_2/U_0=0.1$. 
The inset in (a) shows an enlarged view near the Mott transition point. 
}
\vskip -4mm 
\label{nk}
\end{figure}
%

In Fig.~\ref{nk}(a), the condensate fraction or ${\bm k}=(0,0)$ element 
of the momentum distribution function, 
$
n({\bm k})=\sum_\alpha 
           \langle \hat{a}_{{\bm k},\alpha}^\dag \hat{a}_{{\bm k},\alpha}\rangle, 
$ 
is plotted as a function of $U_0/t$. 
This quantity is finite in the SF phase, but should vanish 
as $1/N_{\rm s}$ in the MI phase.
The value of $|\Psi_{\rm DH}\rangle$ exhibits a sudden drop (for $L\ge 16$) at $U_0=U_{\rm 0c}\sim 22t$ 
and vanishes as $1/N_{\rm s}$ for $U_0>U_{\rm 0c}$ in the inset shown in Fig.~\ref{nk}(a).
$n({\bm k}={\bm 0})/N$ is regarded as an order parameter of the Mott transition here, and so 
this anomaly indicates a first-order SF-MI transition.
In fact, we confirmed that the small-$|{\bm q}|$ behavior of the density 
correlation function $N({\bm q})$ (not shown) changes suddenly from linear 
to quadratic in momentum $|{\bm q}|$ at $U_0=U_{\rm 0c}$. 
These features are substantially identical to those in the spinless 
case on the same D-H binding mechanism\cite{yokoB}.
Next, let us look at the density fluctuation, 
$\sigma_0^2=\langle \hat{n}^2\rangle-\langle \hat{n}\rangle^2$, which is shown 
in Fig.~\ref{nk}(b). 
As is known, the $\sigma_0^2$ of GA completely vanishes at $U_0=U_0^{\rm BR}$, 
which corroborates the view that each site is occupied by exactly one particle 
in the MI phase. 
On the other hand, $\sigma_0^2$ of $|\Psi_{\rm DH}\rangle$ exhibits a small step at 
$U_0=U_{\rm 0c}$ and remains finite for $U_0>U_{\rm 0c}$. 
This finite density fluctuation is reflected in the small 
but finite values of $P(0)\sim P(2)$. 
In the following, we show that the doublon plays a crucial role for the spin structure.

\par 

\begin{figure}
\begin{center}
\vskip 2mm 
\includegraphics[clip,width=8cm]{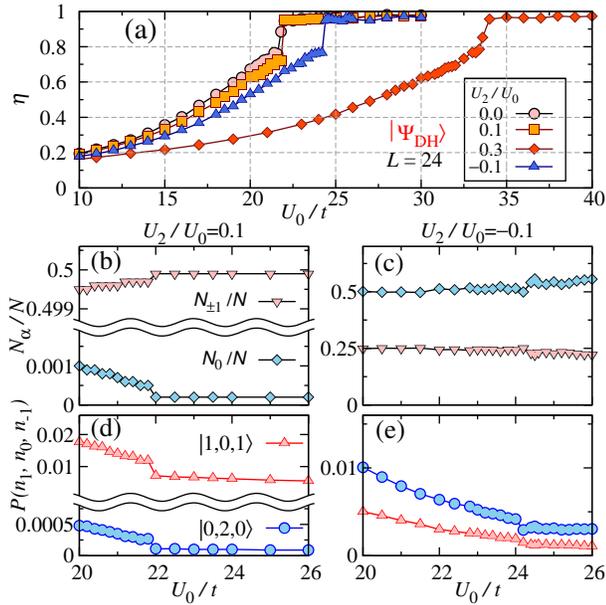}
\end{center}
\vskip -4mm 
\caption{
(Color online)  (a) $U_2/U_0$ dependence of optimized D-H binding parameter $\eta$ as a function of $U_0/t$.
(b),(c)  show the  number fraction of atoms with each spin-component and (d),(e) show the probabilities of doublons with each spin state around $U_{0c}/t$ for $U_2/U_0=\pm 0.1$. 
}
\vskip -4mm 
\label{dh-D}
\end{figure}
%

Now, we turn to the spin-dependent features caused by the $U_{2}$-term 
in ${\cal H}$.
In Fig. \ref{dh-D}(a), the optimized values of the D-H binding parameter 
$\eta$ are plotted to recognize the effects of the $U_{2}$-term on the SF-MI 
transition. 
Recall that $\eta$ controls the strength of the D-H binding between 
nearest-neighbor sites, and $\eta = 1$ means that each doublon is tightly 
bound to an adjacent holon. 
We find that the SF-MI transition point $U_{0c}/t$ shifts to noticeably 
larger values for $U_{2}/U_{0} = 0.3$ and $-0.1$, whereas the shift 
for $U_{2}/U_{0} = 0.1$ is very small.
\par

First, let us consider the difference in the $U_{0c}$ shift between 
for $U_{2}/U_{0} = \pm 0.1$.
Since the system is in the vicinity of the MI phase ($U_{0}\gg t$), 
we may restrict the Fock space to $n_{j} = 0, 1, 2$ at each site; 
actually we confirmed that $n_j\ge 3$ is negligible for $U_{0}/t\gsim 20$. 
To minimize the ground state energy for $U_{2}\neq 0$, not only the number 
of doubly populated ($n=2$) sites but also their spin structures have to 
be optimized. 
To this end, it is convenient to employ the eigenstates of 
$\hat{{\bm S}}_{j}^{2}$, $|S_{j},S_{j}^{z}\rangle\rangle$, where 
$S_{j}$ and $S_{j}^{z}$ indicate the magnitude and $z$-component 
of the total spin at the site-$j$, respectively; $S_{j}$ is even (odd) 
when an even (odd) number of atoms occupy the site-$j$. 
Using $|S_{j},S_{j}^{z}\rangle\rangle$, our basis set 
$|n_{j,1},n_{j,0},n_{j,-1}\rangle$ for $n=2$ is represented as 
\begin{eqnarray}
 |2,0,0\rangle&=&|2,2\rangle\rangle, ~~~
 |1,1,0\rangle= |2,1\rangle\rangle , \nonumber \\
 |0,2,0\rangle &=& \frac{1}{\sqrt{3}}\left( \sqrt{2}|2,0\rangle\rangle + |0,0\rangle\rangle \right), \nonumber \\
 |1,0,1\rangle &=& \frac{1}{\sqrt{3}}\left( |2,0\rangle\rangle - \sqrt{2}|0,0\rangle\rangle \right).
\label{S0S2}
\end{eqnarray}
According to eq. (\ref{S0S2}), the on-site energies of $n=2$ sites 
are calculated as 
$U_{0}+U_{2}$ for $|2,0,0\rangle$ and $|1,1,0\rangle$, 
$U_{0}$ for $|0,2,0\rangle$, and 
$U_{0}-U_{2}$ for $|1,0,1\rangle$.
We also define the probability $p_{\alpha} = N_{\alpha}/N_{s}$ 
($\alpha = -1, 0, 1$), where $N_{\alpha}$ denotes the total number of 
particles of spin $\alpha$; in the present case, $p_{\alpha}$ obeys 
the conditions $0 \leq p_{\alpha} \leq 1$ and $\sum_{\alpha} p_{\alpha} = 1$.
Since we assume $\sum_{j} S_{j}^{z} = 0$, the relation $p_{1} = p_{-1}$ holds, 
resulting in $2p_{1} + p_{0} = 1$. 
Using these formulae, the classical statistical weights for the $n=2$ 
configurations can be calculated as
 $(p_{1})^{2}$ for $|2,0,0\rangle$,  $2(p_{1})^{2}$ for $|1,0,1\rangle$,
 $(p_{0})^{2}$ for $|0,2,0\rangle$,
 and $2p_{0}p_{1}$ for $|1,1,0\rangle$. 
%
For $U_{2} \sim t$, the expectation value of the $U_{2}$-term per site 
is similarly obtained as
\begin{equation}
 E_{2}(p_{1}) \propto -U_{2} p_{1}\left( p_{1} - 1/2 \right) .
\label{E2}
\end{equation}
%
This energy has the minimum value $U_{2}/2$ at $p_{1} = 1/4$ for $U_{2} < 0$,
and $0$ at $p_{1} = 0$ and $1/2$ for $U_{2} > 0$.
Actually, the VMC results in Figs. \ref{dh-D}(b) and \ref{dh-D}(c) show 
$p_{1} = N_{1}/N_{s} \simeq 1/4$ for $U_{2}/U_{0} = -0.1$, 
and $p_{1} \simeq 0.5$ for $U_{2}/U_{0} = 0.1$. 
This causes imbalanced spin population for $n=2$ sites, as shown 
in Figs.~\ref{dh-D}(d) and \ref{dh-D}(e). 
Thus, we find through eq.~(\ref{E2}) that the on-site energies of 
$n=2$ sites are renormalized to $U_{0} + U_{2}/2$ for $U_{2} < 0$, 
while they are not renormalized for $U_{2} > 0$.
Since the onsite energies of $n=2$ sites primarily govern the SF-MI 
transition, this is an appropriate explanation of the difference in the 
$U_{0c}/t$ shifts between for $U_{2}/U_{0} = \pm 0.1$. 
Finally, we point out that the degeneracy in $U_{2}$-energy between 
$p_{1} = 0$ and $0.5$ for $U_{2} > 0$ found in eq. (\ref{E2}) is 
owing to the present spin-1 rotational symmetry, namely, 
$|S^{x} = 0\rangle = ( |S^{z} = 1\rangle + |S^{z} = -1\rangle)/\sqrt{2}$. 
Therefore, we can generate a ground state with $p_{1} \simeq 0$ using VMC 
if we choose a certain initial condition with $p_{0} \gg p_{1}$.
\par

Next, we discuss the difference in the $U_{0c}/t$ shifts between for 
$U_{2}/U_{0} = 0.1$ and $0.3$. 
In the above discussion, we adopted a $classical$ statistical weighting 
[$E_2$ in eq.~(\ref{E2})], which ignores the effect of spin fluctuation 
caused by the $U_{2}$-term, because the spin fluctuation (or singlet 
formation) is suppressed by the particle hopping for small $|U_2|/U_0$'s. 
This is not the case for large $|U_2|/U_0$'s. 
When the $U_2$-term becomes predominant over the hopping term, 
the singlet state 
$|0,0\rangle\rangle=1/\sqrt{3}\left(|0,2,0\rangle-\sqrt{2}|1,0,1\rangle\right)$ 
becomes significant in $n=2$ sites, in order to reduce further the 
$U_{2}$-energy through the spin-exchange processes in the 
$\hat{\bm{S}}^{2}$-term.
In this case, assuming $U_2\gg t$, we ignore the hopping term and 
directly diagonalize the interaction part in ${\cal H}$ using 
$|S_{j},S_{j}^{z}\rangle\rangle$. 
As a result, we find that the on-site energy of $n=2$ sites is renormalized 
toward $U_{0} - 2U_{2}$, which vanishes for $U_{2}/U_{0} = 0.5$.
Thus, the value of $U_{0c}/t$ is bound to grow rapidly as $U_{2}/U_{0}$ 
approaches 0.5, which explains the pronounced shift for $U_{2}/U_{0} = 0.3$ 
in Fig. \ref{dh-D}(a). 
\par

\begin{figure}
\begin{center}
\vskip 2mm 
\includegraphics[clip,width=8cm]{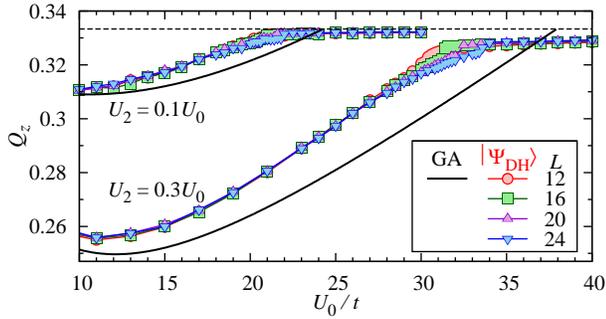}
\end{center}
\vskip -4mm 
\caption{
(Color online) Spin nematic parameter $Q_z$ for two positive values 
of $U_2/U_0$. 
For comparison, the GA result is also plotted, where, 
in fact, $Q_{z}$ cannot be determined for $U_0>U_0^{\rm BR}$. 
}
\vskip -4mm 
\label{snemat}
\end{figure}

\begin{figure}
\vskip 2mm 
\begin{center}
\includegraphics[clip,width=8cm]{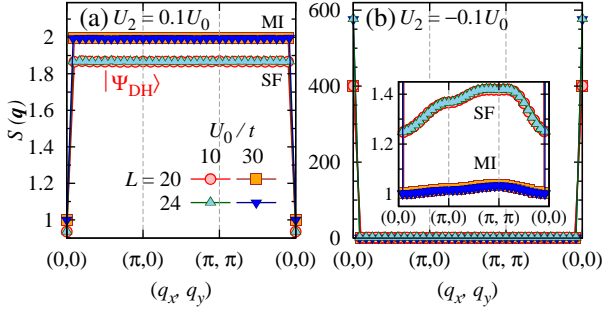}
\end{center}
\vskip -4mm 
\caption{
(Color online) Example of spin structure factor for (a) antiferromagnetic 
coupling ($U_{2}>0)$ and (b) ferromagnetic coupling ($U_{2}<0$).
The phase for $U_0/t=10$ (30) is SF (MI). 
The inset in (b) shows enlarged views of the two phases. 
}
\vskip -4mm 
\label{sq}
\end{figure}

Now, we study the magnetic structures and the spin correlations in the ground state around the SF-MI transition.
For $U_2/U_0>0$, the imbalance of the spin populations found in Fig. \ref{dh-D} (d) suggests that the ground state of the system
exhibits the spin-nematic property.
To detect this with regard to the present case, 
it is useful to study not only spin-correlation functions but also a spin-nematic parameter $Q_{\alpha}$
defined as
\begin{equation}
 Q_{\alpha} = \frac{1}{N_{\rm s}}\sum_{j} 
\left\langle \left(\hat{S}_{j}^{\alpha}\right)^{2} 
                  - \frac{1}{3}\hat{{\bm S}}_{j}^{2} \right\rangle~. 
\label{Q_z}
\end{equation}
%
The maximum value of $Q_{\alpha}$ for a single atom with $S=1$ is 1/3, which is realized in,
e.g., $|1,0,0\rangle$ and $|0,0,1\rangle$,
and also in the eigenstates of $\hat{S}^{x}$ with $S^{x} = 0$, which can be expressed as a coherent
superposition of $|1,0,0\rangle$ and $|0,0,1\rangle$.
Of course, isotropic spin gives $Q_{\alpha} = 0$.
Figure \ref{snemat} shows $Q_{z}$ for two positive values of $U_2/U_0$. 
Reflecting the imbalanced spin population shown in Fig. \ref{dh-D} (d), large values of $Q_{z}$ are
observed in the whole range of $U_{0}/t > 0$, that is, both in the SF and MI phases.
In the MI phase, the single site state is not a coherent superposition of $|1,0,0\rangle$ and $|0,0,1\rangle$
due to the restriction of $\sum_{j}S_{j}^{z} = 0$.
Thus, the spins in the MI phase exhibit a rod-like nematic structure with $Q_{z}\sim 1/3$ and $Q_{x} = Q_{y} \sim -1/6$.
Here, as mentioned in the previous paragraph, the population of doublon $|0,2,0\rangle$  and
thus $N_{0}$ increases as $U_{2}/U_{0}$ increases, which leads to lower $Q_z$ values. 
This is made manifest in that $Q_{z}$ for $U_{2}/U_{0} = 0.3$
is smaller than that for $U_{2}/U_{0} = 0.1$. 
The minima of $Q_{z}$ at $U_0/t\sim 10$ have the same cause, \ie, in the lower $U_0$ region, 
$Q_z$ decreases with increase in $U_0/t$ due to an accompanying increase  of $U_2$, but, for $U_0/t\gsim 10$, 
$Q_z$ enhances as $U_0/t$ increases as a result of decrease of doublon density. 
In the same context, $Q_{z}$ is slightly reduced from the full-moment value 
$1/3$ in the MI phase, where the D-H factor in $|\Psi_{\rm DH}\rangle$ 
induces a particle density fluctuation, in contrast to GA. 
Incidentally, the spin directions remain arbitrary and indefinite 
for $U_2=0$ or $U_0=0$, indicating that the point of $U_2/U_0=0$ is singular. 
In addition, we show the spin structure factor,
$
S({\bm q})=\frac{1}{N_{\rm s}}\sum_{j,\ell}
\langle \hat{{\bm S}}_j\cdot\hat{{\bm S}}_\ell \rangle e^{i{\bm q}\cdot{\bm R}_{j\ell}}, 
$
for $U_2/U_0=0.1$ in Fig.~\ref{sq}(a).
The constant $S({\bm q})$ indicates that there is 
no spin correlation, and the spins take random directions. 
Here, the decrease in $S({\bm q}={\bm 0})$ is reflected in the restriction of $\sum_{j}S_{j}^{z} = 0$.
%
%
Thus, we find a nematic order is realized for $U_2/U_0>0$. 

%
%






\par

%
For $U_2/U_0<0$, $Q_{z}$ becomes negative and its absolute value decreases monotonically 
as $U_0/t$ increases for $U_0<U_{0{\rm c}}$, and is almost constant 
$\sim -1/6$ in the MI phase (not shown), indicating that 
the spins are polarized in the $x$-$y$ plane. 
In Fig.~\ref{sq}(b), $S({\bm q})$ for $U_2/U_0=-0.1$ is plotted in the 
two phases. 
By considering that the value of $S({\bm q}={\bm 0})$ is 
extremely large with almost full moment and diverges proportionally 
to $N_{\rm s}$, a ferromagnetic (FM) long-range order is realized 
in the $x$-$y$ plane. 
The magnitude of the FM moment is almost constant in both the SF and MI phases 
(not shown but expected from Fig.~\ref{sq}(b)). 
The results reported above (for $U_{2} \gtrless 0$) are consistent with those at the weak- (SF, $t \gg U_0 \gg |U_2|$)\cite{yama} 
and strong- (MI, $t \rightarrow 0$)\cite{haradaPRB} interaction limits. 


%
\par
%
In summary, the $S=1$ Bose-Hubbard model [eq.~(\ref{hamil})] on a square lattice 
at unit filling is studied, using a variational Monte Carlo method.
A doublon-holon binding factor $\mathcal{P}_{\rm DH}$, which is the essence of Mott 
transitions, not only improves the variational energy considerably 
upon the GA, but enables us to study the details of 
the spin structure directly 
even in the MI phase without resorting to an effective 
spin Hamiltonian. 
For $U_2>0$ ($<0$), a spin nematic (ferromagnetic) phase is stabilized 
from SF to MI phases. 
The present results broadly support the phase diagrams\cite{krut,tsuchiya,kimura} 
and spin structure\cite{yama,haradaPRB} as regards $S=1$ BHM proposed in previous studies.
\par

\begin{acknowledgment}
Some of the numerical computations were carried out at the Yukawa Institute 
Computer Facility and at the Cyberscience Center, Tohoku University.
This work is supported by a Grant-in-Aid for Scientific Research (C) 
and also by the Next Generation Supercomputing Project, Nanoscience Program, 
from MEXT 
of Japan.
\end{acknowledgment}
\par 



\end{document}